\documentclass[showpacs,amsmath,amssymb,aps,twocolumn,longbibliography,pra]{revtex4-2}
\usepackage{graphicx}
\usepackage[colorlinks=true,citecolor=blue,linkcolor=magenta]{hyperref}
\usepackage[usenames]{color}
\usepackage{relsize}
\usepackage[english]{babel}
\usepackage{enumerate}
\usepackage{bbm}

\newcommand{\ket}[1]{\left| #1 \right>} 
\newcommand{\bra}[1]{\left< #1 \right|} 

\newcommand {\grsim} {\ {\raise-.5ex\hbox{$\buildrel>\over\sim$}}\ }
\newcommand {\lessim} {\ {\raise-.5ex\hbox{$\buildrel<\over\sim$}}\ } 
\newcommand {\ii} {i}

\usepackage{xcolor}
\usepackage{siunitx}

\newcommand{\RN}[1]{%
  \textup{\uppercase\expandafter{\romannumeral#1}}%
}

\usepackage{xr}
\newcommand{\nocontentsline}[3]{}
\newcommand{\tocless}[2]{\bgroup\let\addcontentsline=\nocontentsline#1{#2}\egroup}

\begin{document}

\title{Benchmarking multi-qubit gates - II: Computational aspects }

\author{Bharath Hebbe Madhusudhana }
\author{}

\affiliation{$^{1}$\,Fakult\"at f\"ur Physik, Ludwig-Maximilians-Universit\"at M\"unchen, Schellingstra{\ss}e 4, 80799 M\"unchen, Germany}
\affiliation{$^{2}$\,Munich Center for Quantum Science and Technology (MCQST), Schellingstr. 4, 80799 M\"unchen, Germany}
\affiliation{$^{3}$\,Max-Planck-Institut f\"ur Quantenoptik, Hans-Kopfermann-Stra{\ss}e 1, 85748 Garching, Germany}


\begin{abstract} 
An important step in developing multi-qubit gates is to construct efficient benchmarking protocols for them. In our previous paper (arXiv: 2210.04330), we  developed metrological protocols to measure the reduced Choi matrix  i.e., the completely positive (CP) maps induced on a subset S of the qubits, by the multi-qubit gate. Here, we show a set of classically verifiable properties that the Choi matrix satisfies if it is a reduction of a multi-qubit unitary and use them to develop benchmarks. We identify three types of errors that affect the implementation of a multi-qubit unitary, based on their mathematical properties and physical origin. Although a target multi-qubit gate is a unitary operator, errors turn it into a general completely positive (CP) map.  Errors due to coupling to a thermal bath result in the multi-qubit gate being CP-divisible (Markovian), deviating from a unitary. The reduced Choi matrix of a multi-qubit gate has a property known as \textit{double stochasticity}, which is violated in the presence of Markovian errors. We construct a benchmark using double-stochasticity violation and show that it is sensitive to coupling to any thermal bath at a finite temperature.  Further, errors due to shot-to-shot fluctuations result in a non-markovian, i.e., \textit{CP-indivisible quantum process}. We prove a new property, which we call the \textit{rank property} of the reduced Choi matrix, the violation of which implies a CP-indivisible error. A third category of errors comes from systematics in the implementation of a multi-qubit gate, resulting in no deviation from unitarity.. We refer to this as unitary errors. This corresponds to the most challenging type of error to benchmark. We develop a partial-benchmarking protocol for such errors using symmetries of the multi-qubit gate being applied. 
\end{abstract}

\maketitle

\textbf{}

\tocless\section{Introduction}

The gate set of most state-of-the-art quantum computers consists of one and two-qubit gates, using which one can, in principle,  construct arbitrary unitary operations on the full system of $N$ qubits in the quantum computer. One can expand this gate set by developing multi-qubit gates.  The latter is a unitary acting on $N$ qubits, $U\in SU(2^N)$, which entangles more than two (presumably all) qubits.  It can be produced by time evolution under a many-qubit Hamiltonian $H$, $U=e^{-\ii Ht}$.  While the theory developed in this paper is aimed at such multi-qubit gates, most of it is also applicable for multi-qubit operations generated by a circuit consisting of one and two-qubit gates. 

Having multi-qubit gates in the gate set offers an advantage in circuit optimization. That is, there will be multiple ways to implement a given algorithm and that helps us optimize the error. As a result, there has been some interest in the development of multi-qubit gates recently~\cite{article_multi_qubit, Martinez_2016, PRXQuantum.2.040348}. Multi-qubit gates also have applications in \textit{quantum certified approximations}, i.e., benchmarking the performance of a new classical approximation ansatz for many-body systems using an analog quantum computer~\cite{PRXQuantum.2.040325}.  Moreover, a classical neural network can be trained on the data produced by multi-qubit gates~\cite{Huang_2022, Huang_2022_2}.  Other applications of multi-qubit gates have been studied recently~\cite{https://doi.org/10.48550/arxiv.2210.02936, Qsim_review}.  A \textit{digital-analog quantum computer},  is a device where universal quantum control is achieved using a gate set consisting of single-qubit gates and a few multi-qubit gates~\cite{PhysRevA.101.022305, PhysRevResearch.2.013012, Yu_2022}.

In the previous paper~\cite{https://doi.org/10.48550/arxiv.2210.04330} we identified two challenges in benchmarking multi-qubit gates: (i) metrological challenge, referring to the unfeasibility of a complete process tomography~\cite{Chuang_1997} and (ii) computational challenge, which involves finding benchmarks that can be computed efficiently on a classical computer. We addressed the metrological challenges using the reduced Choi matrix corresponding to a subset $S\subset \{1, 2, \cdots, N\}$ of qubits. The latter represents the quantum channel corresponding to $S$ induced by the multi-qubit gate acting on the $N$ qubits.  We developed reduced process tomography protocols to measure this reduced Choi matrix. In particular, we used entanglement and information scrambling to speed up the convergence rate of the sampling error in the reduced process tomography. 

We address the computational challenge in this paper. That is, given an experimentally measured reduced Choi matrix, we consider the question: what are the figures of merit/benchmarks which estimate the errors in the gate and which can be computed efficiently on a classical computer using the reduced Choi matrix? 

Besides the practical utility of estimating errors, some of the benchmarks are also important at a conceptual level, in deciding whether a quantum advantage has been achieved.  The choice of the benchmarks used to establish quantum advantage is a hotly debated topic~\cite{https://doi.org/10.48550/arxiv.2109.11525, https://doi.org/10.48550/arxiv.2103.03074}. The cross-entropy benchmark was among the first figures of merit used experimentally to establish a quantum advantage~\cite{Sycamore_2019}.  One of the earliest theoretical benchmarks is the average fidelity~\cite{NIELSEN2002249} used in randomized benchmarking~\cite{PRXQuantum.3.020357, Emerson_2005} and Clifford gate benchmarking~\cite{PhysRevA.77.012307, PhysRevLett.106.180504, Proctor_2021}.  Recently benchmarks based on the expected statistical properties of the output state have been developed.  For instance, one expects the overlaps of the output state on the basis to follow a  Porter-Thomas distribution, when the unitary evolution has strong information scrambling~\cite{arxiv.2103.03536, arxiv.2103.03535}.    

In contrast to these results, we use the mathematical properties of completely positive maps that can be written as unitary partial traces to develop efficient benchmarks which also characterize most of the physically realistic errors in the system. We begin with a summary of our results.\\

\begin{figure}
\includegraphics[scale=0.35]{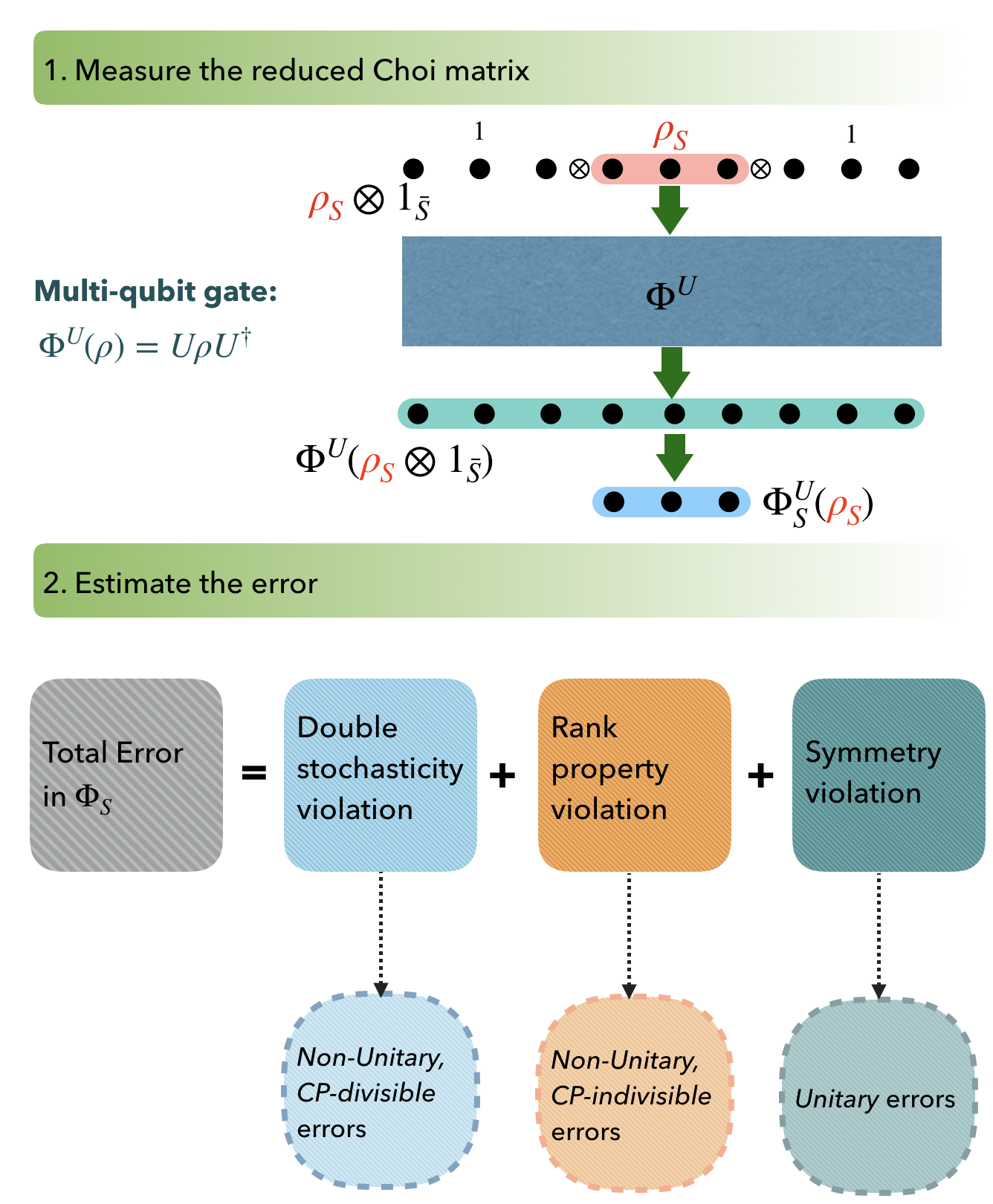}
\caption{\textbf{Benchmarking multi-qubit gates:} Errors in multi-qubit gates can be estimated using the reduced Choi matrix. The metrological aspects of measuring the latter were considered in the previous work. In this work, we show how to estimate the error using the reduced Choi matrix. In particular, we identify three \textit{physically }distinct sources of error that leave three \textit{mathematically }distinct signatures in the reduced Choi matrix. We show how to utilize the mathematical signatures to estimate the contribution of each of the three errors.  }\label{Fig1}
\end{figure}

\tocless\section{Results}
Let U be a unitary acting on N qubits and $\Phi_U$ be the corresponding completely positive map. Benchmarking an experimental realization $\Phi$ against $\Phi_U$ fundamentally relies on the properties of $\Phi_U$ that can be efficiently computed on a classical computer. For instance, $\Phi_U$ represents a unitary resulting in certain constraints on it. Moreover, if $U$ has known symmetries, that results in additional constraints on $\Phi_U$. Here, we show how these properties can be used to develop benchmarking protocols. It must be noted that one of the consequences of quantum advantage is that $\Phi_U$ itself cannot be computed on a classical computer, in general. One of the most common forms of error is systematic, resulting in a unitary process $\Phi$, but a different one from $\Phi_U$. We refer to this as a unitary error. A coupling to the environment, for instance, a light-assisted decay, would result in a non-unitary $\Phi$. Moreover, random, shot-to-shot fluctuations of the control parameters would also result in a non-unitary $\Phi$. We refer to the latter two as non-unitary errors. 

One can, in principle, check for non-unitary errors, if $\Phi$ is fully characterized, by checking for unitarity. However, a complete characterization of $\Phi$ is not scalable. Moreover, checking for unitarity is also not an efficient process. In the previous work, we introduced a reduced process tomography, and developed metrology protocols for it. Using these protocols, we can characterize $\Phi^S$, the process restricted to a subset $S$ of the $N$ qubits. The obvious question is, to what extent can we check for unitarity of $\Phi$, using $\Phi^S$? On the surface, the Stinespring dilation theorem appears to suggest that \textit{every} $\Phi^S$ can be dilated into a unitary map and therefore, it would not be possible to check for unitarity using $\Phi_S$. However, the Stinespring dilation requires the remaining qubits in $\bar{S}$($=\{1, \cdots, N\} - S$) to be in a pure state, whereas, in our definition of the reduced process, they are in the maximally mixed state (se supplementary information for details). We show that it is indeed possible to show some necessary criteria on $\Phi^S$ for $\Phi$ to be unitary. For instance, a unitary operation maps $\mathbbm{1}$ to $\mathbbm{1}$ and therefore, the corresponding reduced process also satisfies $\Phi^S(\mathbbm{1}) = \mathbbm{1}$. Together with trace preservation, this property is called \textit{double stochasticity}~\cite{LANDAU1993107}.  In section~\ref{NU_markovian}, we show that double stochasticity violation can be used to benchmark some of the non-unitary errors. In particular, we show that if the qubits are coupled to a thermal bath, the double stochasticity of $\Phi^S$ is necessarily violated.  That is, coupling to a thermal bath can always be characterized by double stochasticity violation.

There is a fundamental difference between non-unitary errors due to coupling to a thermal bath and due to random shot-to-shot fluctuations. The former is \textit{Markovian} or \textit{P-divisible}, while the latter is not.  We show that most physically relevant P-divisible errors can be characterized via double stochasticity violation. However, the shot-to-shot fluctuations, which are the most relevant non-Markovian errors cannot be characterized using double stochasticity violation --- the process $\Phi^S$ remains doubly stochastic even in the presence of such errors.  In order to tackle this problem, we show a new property called the \textit{rank property}, satisfied by the reduced process $\Phi^S$ if $\Phi$ is unitary.  We prove this in Theorem 2 in section~\ref{NU_Nmarkovian} We also show, using numerical simulations, that most random fluctuations result in a violation of the rank property and therefore, can be characterized. 

The above two techniques, using double stochasticity and the rank property, are useful only to characterize non-unitary errors. Unitary errors, caused by systematics in the experiment result in a unitary $\Phi$, different from $\Phi_U$ and it will therefore maintain both double stochasticity and the rank property.  We can tackle this problem, at least partially, using the known symmetries of $U$. For instance, consider $U=e^{-i Ht}$, generated by a multi-qubit Hamiltonian $H$.  This unitary commutes with $H$ and therefore $\Phi_U$ conserves $\langle H\rangle, \langle H^2\rangle, \cdots$. A measurement of the time variation of $\langle H\rangle$ can be used to characterize some of the unitary errors in $\Phi$.  For example, consider an XXZ Hamiltonian, $H = J \sum_i \sigma_{x, i}\sigma_{x, i+1} + \sum h_i \sigma_{z, i}$. This is commonly used in trapped neutral atomic systems.  The expectation value $\langle H \rangle$ can be constructed by measuring local observables $\langle \sigma_{z, i}\rangle$ and nearest neighbour correlations $\langle \sigma_{x, i}\sigma_{x, i+1}\rangle$.  In section~\ref{Unitary} we prove an inequality to estimate the unitary error using the observed deviation of $\langle H \rangle$.  More generally, we show that any known symmetry of $U$, i.e., an observable $X$ that can be measured and is known to commute with $U$ can be used to estimate the unitary errors. In particular, the recently developed ideas on Fragmentation~\cite{PhysRevLett.130.010201} can be used to identify conserved quantities that can be used for benchmarking.  

This paper is organized as follows. In section~\ref{error_types}, we describe the classification of errors in quantum processes into unitary, non-unitary, Markovian, and non-Markovian.  In section~\ref{NU_markovian}, we present our results on using double stochasticity to benchmark non-unitary, P-divisible errors. Following, in section~\ref{NU_Nmarkovian}, we state and prove the rank property and present our results on benchmarking non-unitary, non-P-divisible errors. In  section~\ref{Unitary}, we show how to benchmark unitary errors using symmetries.  We outline experimental protocols that can be implemented in state-of-the-art devices in section~\ref{Expt_protocols}. Finally, in section~\ref{conclusions}, we conclude with a few remarks on possible error mitigation strategies. \\

\tocless\section{Errors in Quantum Processes}\label{error_types}

We consider the application of a quantum gate on $N$ qubits as a time-dependent completely positive map $\Phi_t$ where $\Phi_{t=0}=\text{Id}$ is the identity map and $\Phi_{t=1}$ is the desired gate.  The most desirable quantum gate is a unitary trajectory, i.e., $\Phi_t$ is unitary for each $t$ and $\Phi_{t=1} = \Phi_U$ where $U$ is the target gate.  However,  such maps form a zero-measure subset of the set of all time-dependent operations $\Phi_t$.  Most errors produce a non-unitary $\Phi_t$.  Below we present a convenient classification of the errors, based on their physical origin and the  mathematical properties of the resulting CP map.\\

\begin{figure*}
\includegraphics[scale=0.42]{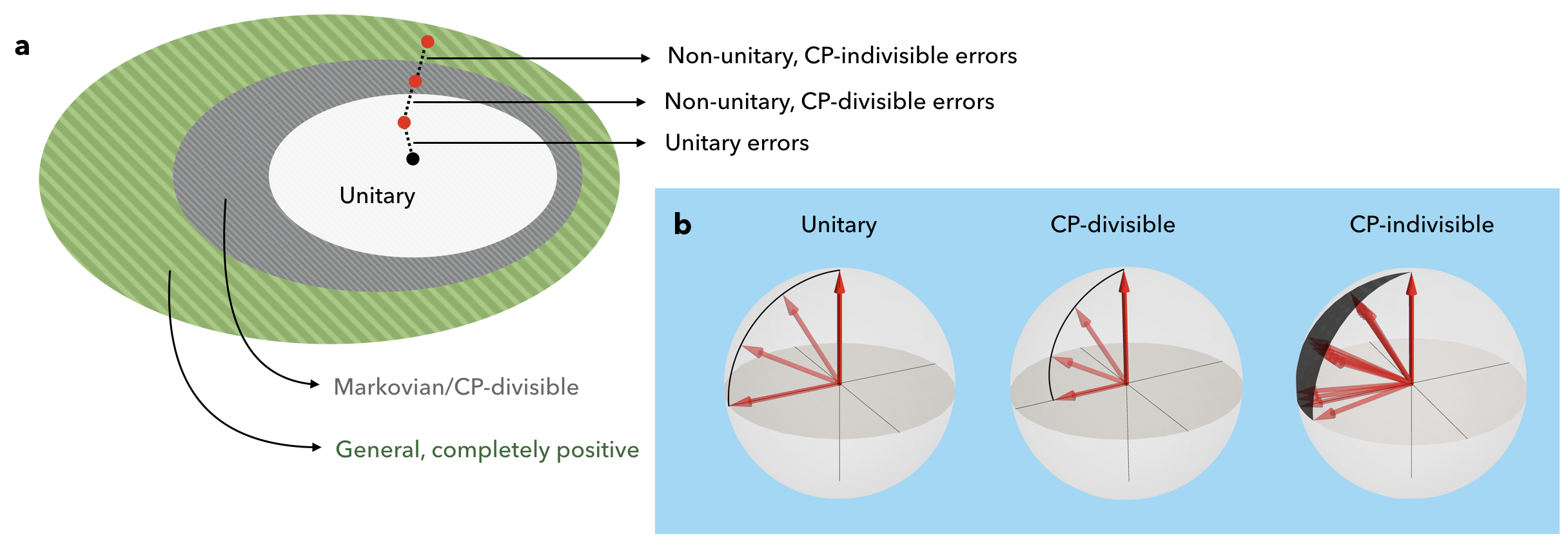}
\caption{\textbf{Errors in quantum channels:} \textbf{a} Quantum processes can be classified into three nested subsets: Unitary, CP-divisible, and completely positive (see text for definitions).  Accordingly, the errors in a target Unitary operation can be classified into Unitary errors, which maintain the process within the subset of Unitary processes, Non-unitary CP-divisible errors, which move the process into the CP-divisible subset and Non-unitary, CP-indivisible errors, which move it to the general of CP maps. The three types of errors have different physical origins.  \textbf{b} examples of the three types of processes for a single qubit system. A rotation is a unitary operation; a rotation along with a Bloch-Redfield type decay is a CP-divisible process and a rotation with a fluctuating axis and rate is a general CP-indivisible process.   }\label{Fig2}
\end{figure*}

\tocless\subsection{Classification of the errors}

A common form of error is coupling to the environment, where $\Phi_t =\mathcal T e^{\mathcal{L} t}$ is the solution to a Lindblad master equation. Here, $\mathcal T$ is time-ordering and $\mathcal L$ is:
\begin{equation}\label{lindblad}
\mathcal L (\rho) = -i[H(t), \rho] + \sum_j L_j\rho L_j^\dagger -\frac{1}{2}\{\rho, L_j^\dagger L_j\}
\end{equation}
$H(t)$ is the Hamiltonian and $L_j$ are the jump operators representing the coupling to environment~\cite{doi:10.1063/1.5115323}. Such a quantum operation is non-unitary and is a very general description of quantum processes. However, it is not the most general --- there are processes occurring in the lab that can't be described by the above equation.  The main reason is, this process is CP-divisible(i.e, completely positive divisible), or Markovian~\cite{Chruciski_2012}. That is, it satisfies 
\begin{equation}
\Phi_{t}=\Phi_{t-t'}\circ\Phi_{t'}
\end{equation}
for $t'\leq t$.  An elementary example is the Bloch-Redfield model, describing the decay of a qubit from $\ket{1}$ to $\ket{0}$ (see ref.~\cite{suppmat}).  It follows from the GKLS theorem that every CP-divisible process can be described by a master equation of the form Eq.~\ref{lindblad}.  The set of CP-divisible processes forms a bigger set that includes unitary processes Fig.~\ref{Fig2}. 

A common form of error that moves the quantum process outside the CP-divisible set is decoherence, resulting from shot-to-shot fluctuations~\cite{Rivas_2014, PhysRevA.97.012127}.   Consider, for example, a unitary process generated by a time-independent Hamiltonian $H$: $\Phi_t : \rho \mapsto e^{-iHt}\rho e^{i Ht}$. If the implementation of this process involves a fluctuation of $H$ between experimental shots, described by some measure $dH$ (possibly Gaussian), the resulting process is $\Phi_t: \rho \mapsto \int dH e^{-iHt}\rho e^{i Ht}$. This process is in general not divisible
\begin{equation}
\begin{split}
\Phi_{t+t'} :&  \rho \mapsto \int dH e^{-iH(t+t')}\rho e^{i H(t+t')}\\
 \neq \Phi_{t}\circ \Phi_{t'}:&  \rho \mapsto  \int dH dH' e^{-iHt}e^{-iH't'}\rho e^{iH't'}e^{i Ht}
\end{split}
\end{equation}
A common experimental situation is a Gaussian fluctuation of applied fields (see ref.~\cite{suppmat} for an explicit example of a non-CP-divisible process).  In general, one may have errors coming from both coupling to the environment and shot-to-shot fluctuations.  

In the next two sections, we show that the distinctive nature of the above two types of errors (CP-divisible non-unitary errors and CP-indivisible non-unitary errors) can be used to characterize them in an experiment. However, a third category of errors arising from systematic deviations of the applied Hamiltonian from the target maintains the unitarity of the process and therefore more challenging to characterize.   We refer to these as unitary errors.  In the presence of only unitary errors,  the implementation of a target unitary gate $\Phi_U$ would be a different unitary $\Phi_{U'}$.  We return to address the problem of characterizing unitary errors in section~\ref{Unitary}.

Before proceeding with the benchmarking of non-unitary, CP-divisible errors, we address the question of what is the appropriate quantification of the error, which will be necessary for the benchmarking. \\

\tocless\subsection{Quantification of the error}

Let $\Phi$ and $\Phi'$ be two maps. Our goal here is to quantify the difference between them in a meaningful way. One can consider the algebraic difference between the corresponding Choi matrices $\epsilon = \rho^\Phi - \rho^{\Phi'}$.  The physically relevant aspect of this difference is in the expectation value of an observable $O$ with respect to $\Phi(\rho)$ and $\Phi'(\rho)$: $\Delta \langle O \rangle_\rho =\langle O\rangle_{\Phi(\rho)} - \langle O\rangle_{\Phi'(\rho)} = \text{Tr}(\epsilon O\otimes \rho)$.  This represents the measurable  difference in the observable expectation value when starting with the same initial state $\rho$ through the channels $\Phi$ and $\Phi'$.  One can use the below inequality
\begin{equation*}
|\text{Tr}(\epsilon O\otimes \rho)| \leq ||O||_2 ||\epsilon||_2
\end{equation*}
to define $||\epsilon||_2$ as the measure of the error. Here, $||\cdot ||_2$ represents the Schatten 2-norm. Note that $||\rho||_2 \leq 1$.  However, this inequality is far from being tight because $\epsilon$ is unlikely to be a tensor product.  One can also show a different inequality,
\begin{equation}
|\text{Tr}(\epsilon O\otimes \rho)| \leq ||O||_2 \sigma_{\max}(\Phi-\Phi')
\end{equation}
Here, $\sigma_{\max}(\Phi-\Phi')$ is the maximum singular value of the map $\Phi-\Phi'$. Note that this is \textit{not} the maximum singular value of $\epsilon$, treated as a matrix.  This inequality is also no tight, in general. However, it is tighter than the previous one.  We use $\sigma_{\max}(\Phi-\Phi')$ as a measure of the error in the quantum process in the rest of the paper.  See ref.~\cite{suppmat} for an example illustrating the measure. \\

\tocless\section{Non-unitary, CP-divisible errors}\label{NU_markovian}

In this section, we will develop a benchmarking protocol to estimate errors due to non-unitary, CP-divisible processes, such as coupling to the environment. Let $\Phi_U$ be the CP map corresponding to a target unitary $U$ acting on $N$ qubits and $\Phi$ be the CP map corresponding to its experimental implementation. For a subset $S\subset \{1, 2, \cdots, N\}$, we assume that $\Phi^S$ has been measured experimentally and is to be compared with $\Phi_U^S$, the target reduced process corresponding to $S$. We assume that $\Phi_U^S$ is \textit{unknown}, or cannot be computed classically. The goal is to estimate the error $\sigma_{\max}(\Phi_U^S -\Phi^S)$, using only $\Phi^S$. 

By definition,  $\Phi_U^S(\rho^S) = \text{Tr}_{\bar{S}}[U\rho^S\otimes \mathbbm{1}^{\bar{S}}U^{\dagger}]$, it follows that $\Phi_U^S(\mathbbm{1}^S)=\mathbbm{1}^S$.  This should hold for any unitary  $U$. Moreover,  the map is trace preserving $\text{Tr}\Phi_U^S(\rho^S) = \text{Tr}_{S}\text{Tr}_{\bar{S}}[U\rho^S\otimes \mathbbm{1}^{\bar{S}}U^{\dagger}]= \text{Tr}(\rho^S)$. These two properties together constitute double stochasticity~\cite{LANDAU1993107, GOWDA201740}.\\

\noindent \textbf{Definition 1:} A map $\Phi$ is doubly stochastic iff it is trace preserving, $\text{Tr}[\Phi(\rho)]=\text{Tr}(\rho)$, and identity preserving, $\Phi(\mathbbm{1})=\mathbbm{1}$. 

It follows that if $\Phi$ is unitary, then \textit{every} reduction of it, $\Phi^S$ is doubly stochastic.  The below lemma establishes the double stochasticity as a criterion on the Choi matrix:\\

\begin{figure}
\includegraphics[scale=0.4]{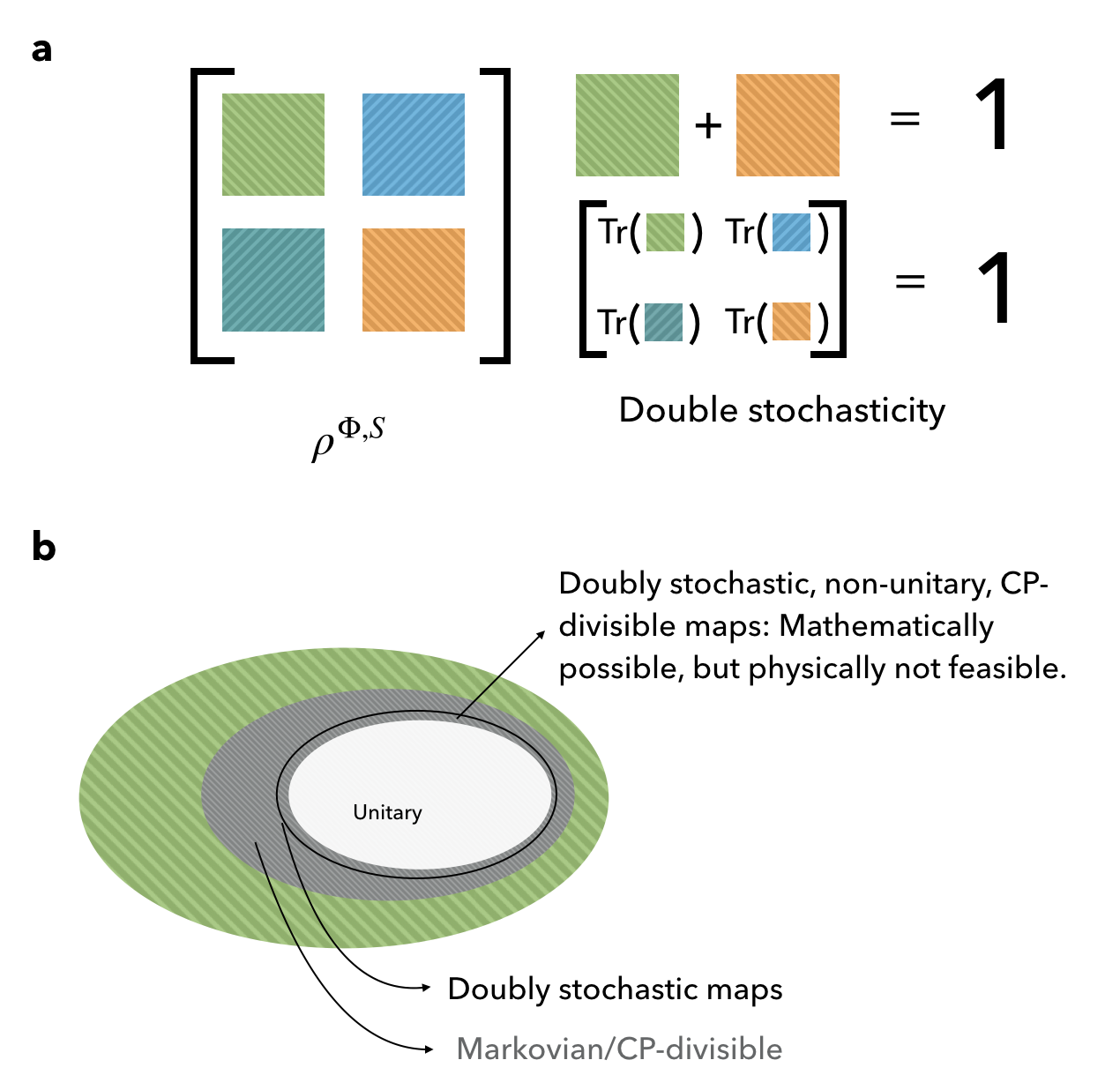}
\caption{\textbf{Benchmarking non-unitary, CP-divisible errors:} \textbf{a.} Illustrates the definition of double stochasticity. A Choi matrix $\rho^{\Phi, S}$ is doubly stochastics iff both of its partial traces are equal to identity.  \textbf{b.} shows a Venn diagram illustrating doubly stochastic maps. Violation of double stochasticity strictly implies CP-divisible, non-unitary errors. Conversely, the presence of physically realistic non-unitary, CP-divisible errors implies a violation of double stochasticity, making the latter a good measure of non-unitary, CP-divisible errors. }\label{Fig3}
\end{figure}

\noindent \textbf{Lemma 1:} If $S$ is coupled only to $\bar{S}$, then $\Phi^S$ is doubly stochastic. Moreover, the corresponding Choi matrix $\rho^{S, \Phi}$ satisfies $\text{Tr}_{\text{in}}(\rho^{S, \Phi})=\text{Tr}_{\text{out}}(\rho^{S, \Phi})=\mathbbm{1}$. 

\noindent\textbf{Proof:} Let $d=2^{|S|}$ be the dimension of the Hilbert space of $S$. $\rho^{S, \Phi}$ is a $d^2\times d^2$ matrix, defined as $\rho^{S, \Phi}_{ij; kl} = \bra{j}\Phi^S(\ket{i}\bra{k})\ket{l}$, where $\ket{i}, i =1, \cdots d$ form a basis. One can interpret $i, k$ as the "input" qubit states and $j, l$ as the "output" qubit states.  $\text{Tr}_{\text{in}}$ and $\text{Tr}_{\text{out}}$ above refer to the partial trace w.r.t to the input and output qubit states. 

The double stochasticity follows from the arguments presented before.  Moreover, $\Phi^S(\mathbbm{1}) = \sum_{ijl}\ket{j}\bra{l}\Phi^S(\ket{i}\bra{i})= \text{Tr}_{\text{in}}\rho^{S, \Phi}$. Thus, it follows that $ \text{Tr}_{in}\rho^{S, \Phi}=\mathbbm{1}$. Finally, from the trace preservation and $\text{Tr}[\Phi^{S, \Phi}(\rho)] = \sum_{ikj}\bra{j}\Phi^S(\ket{i}\bra{k})\ket{j}\rho_{ik} = \text{Tr}[\text{Tr}_\text{out}(\rho^{S, \Phi})\rho]$, it follows that $\text{Tr}[\text{Tr}_\text{out}(\rho^{S, \Phi})\rho]= \text{Tr}(\rho) \forall \rho$. Thus, $\text{Tr}_\text{out}(\rho^{S, \Phi})=\mathbbm{1}$.

This lemma shows a necessary condition for unitarity. Violation of double stochasticity of $\Phi^S$ indicates non-unitary errors in the process.  The two obvious follow-up questions are: (i) to what extent is the converse true; that is, what is the class of non-unitary errors that will necessarily result in double stochasticity violation? and (ii) To what extent can we quantitatively estimate the error $\sigma_{\max}(\Phi^{S}-\Phi^S_U)$ using double stochasticity violation? We return to the former question after addressing the latter. \\

A simple estimate is the lower bound given by $\sigma_{\max}(\Phi^{S}-\Phi^S_U)\geq \frac{||\mathbbm{1}-\Phi^S(\mathbbm{1})||_2}{\sqrt{d}}$. This is the most straightforward bound to compute and follows from the fact that $\sigma_{\max}(\Phi^{S}-\Phi^S_U) \geq \frac{||\Phi^{S}(\mathbbm{1})-\Phi^S_U(\mathbbm{1})||_2}{||\mathbbm{1}||_2}$ We will show, below, that this in indeed also very useful, despite its simplicity. Another obvious lower bound is,
\begin{equation*}
 \min_{\Phi}\{\sigma_{\max}(\Phi^S-\Phi):\ \Phi \text{ CP, doubly stochsatic}\}
\end{equation*}
One can construct a physically more meaningful measure or error, by imposing a structure on the operators $\Phi$ above.  For any given CP, doubly stochastic map $\Phi$, one can find a map $\chi$ such that $\Phi^S =\chi \circ \Phi$. However, one cannot guarantee that $\chi$ represents a physical process, i.e., that it is CP or trace-preserving.  A physically reasonable assumption is that $\Phi^S =\chi \circ \Phi_U^S$, where $\chi$ represents the physical process that causes the error and is therefore CP and trace-preserving. We can now minimize $\sigma_{\max}(\Phi^S - \Phi)$ where in addition to $\Phi$ being CP, doubly stochastic, there exists a CP, trace-preserving map $\chi$ such that $\Phi^S =\chi\circ \Phi$.  
\begin{equation}\label{eps}
\begin{split}
 \epsilon=& \min_{\Phi}\{\sigma_{\max}(\Phi^S-\Phi):\ \Phi, \text{ CP, doubly stochsatic and }\\
 & \chi\circ \Phi=\Phi^S\ \text{ where }\chi\text{ CP, trace preserving} \}
 \end{split}
\end{equation}
This, however, is a more complicated optimization to perform.  Below, we show that $\epsilon$ is well-approximated by the measure $\sqrt{d}||\mathbbm{1}-\Phi^S(\mathbbm{1})||_2$. 

\noindent\textbf{Theorem 1} 
\begin{equation}
||\mathbbm{1}-\Phi^S(\mathbbm{1})||_2 \geq \epsilon \geq \frac{||\mathbbm{1}-\Phi^S(\mathbbm{1})||_2}{\sqrt{d}}\\
\end{equation}
\textbf{Proof:} The second inequality follows from the observation 
$$\sigma_{\max}(\Phi^S-\Phi) \geq \frac{||\Phi^S(\mathbbm{1}) -\Phi(\mathbbm{1})||_2}{||\mathbbm{1}||_2}= \frac{||\mathbbm{1}-\Phi^S(\mathbbm{1})||_2}{\sqrt{d}}$$. 

To prove the first inequality, note that $\Phi^S(\mathbbm{S})\subseteq\mathbbm{S}$, where $\mathbbm{S}$ is the set of PSD matrices. This follows from the complete positivity of $\Phi^S$. Moreover, every point in the interior of $\mathbbm{S}$ is mapped to a point in the interior of $\mathbbm{S}$. Thus, if $\Phi^S(\mathbbm{1})\neq \mathbbm{1}$, then $\Phi^S(\mathbbm{S})\neq \mathbbm{S}$. Let us consider a CP map $\chi$ such that $\chi(\mathbbm{1})=\Phi^S(\mathbbm{1})$ and $\chi(\mathbbm{S})\supseteq \Phi^S(\mathbbm{S})$. It then follows that the map $\chi^{-1}\circ\Phi^S$ is a member of the set considered in Eq.~\ref{eps}. Thus,
$$
\epsilon \leq \sigma_{\max}(\chi^{-1}\circ\Phi^S -  \Phi^S) \leq \sigma_{\max}(\chi^{-1}-Id)\cdot \sigma_{\max}(\Phi^S)
$$
This holds for all $\chi$ satisfying the said conditions.  Thus, the problem boils down to minimizing $\sigma_{\max}(\chi^{-1}-Id)$.  It follows, based on the conditions, that this minima is $O(||\Phi^S(\mathbbm{1})-\mathbbm{1}||_2)$.  Thus the result follows $\blacksquare$

We now turn to the first question: what is the class of non-unitary errors that will necessarily result in double stochasticity violation? The classic theory of open quantum systems~\cite{markov_1, Rivas_2010, Markovian_2} already provides us with the answer:

\noindent \textbf{Lemma 2:} If $S$ is coupled to a third system $E$ (besides $\bar{S}$), which is in an ergodic state with a finite temperature $T<\infty$, then $\Phi^S$ necessarily violates double stochasticity. 

\textbf{Proof:} From the theory of open system dynamics, it follows that,  in the limit of infinite time,  all states are mapped to $\rho_{th}(T)$, i.e., the thermal state with temperature $T$. Thus  in particular, $\Phi^S(\mathbbm{1}) \rightarrow \rho_{th}(T) \implies \Phi^S(\mathbbm{1})\neq\mathbbm{1}$. That is, the system has one and only one fixed point and that is $\rho_{th}(T)\neq \mathbbm{1}$ when $T<\infty$. Thus, it necessarily violates double stochasticity $\blacksquare$

To summarize the results in this section, double stochasticity is a strictly necessary condition and a sufficient condition under physically reasonable assumptions for the absence of non-unitary CP-divisible errors.  The cases where a reduced Choi matrix does not violate double stochasticity, but nevertheless contains non-unitary, CP-divisible errors represent mathematically possible, but physically unfeasible error processes (see Fig.~\ref{Fig3}). Moreover, Theorem 1 provides a simple estimate to the double stochasticity violation. Thus, double stochasticity violation is, for practical purposes, a very powerful measure of non-unitary, CP-divisible error.   \\

\tocless\section{Non-unitary,  CP-indivisible errors}\label{NU_Nmarkovian}
While double stochasticity is violated for most CP-divisible errors,  it is preserved under a large class of non-unitary errors. Consider a general shot-to-shot fluctuation of a target unitary $U$. The resulting process is non-unitary, given by $\Phi (\rho) = \int dU U\rho U^{\dagger}$. Note that this averaging necessarily preserves the double stochasticity of $\Phi$ and all its reduced Choi matrices. This process, as argued in section~\ref{error_types}, is CP-indivisible and represents one of the most commonly occurring forms of error.  One cannot detect such errors via double stochasticity violation. In this section, we prove a new property of the reduced Choi matrix, which we refer to as the \textit{rank property} and show that it's violation can be used to quantify and benchmark non-unitary, CP-indivisible errors. 

The task is to determine if $\rho^{\Phi, S}$ is the partial trace of a bigger Choi matrix, which corresponds to a unitary operation. An apparent hurdle to this task is the Stinespring dilation theorem, which says that every Choi matrix can be \textit{dilated} into a unitary.  However, on closer inspection, we find that the Stinespring dilation of $\rho^{\Phi, S}$ does not produce a Choi matrix whose partial trace is $\rho^{\Phi, S}$. Instead, it produces a unitary operation $\Phi(\sigma)=U\sigma U^{\dagger}$ and a pure state $\ket{\eta}$ such that
$$
\Phi_S(\rho) = \text{Tr}_2U\rho\otimes \ket{\eta}\bra{\eta}U^{\dagger}
$$
Here, $\ket{\eta}\bra{\eta}$ is strictly pure; whereas, what we need is a unitary operation $U$ that satisfies
$$
\Phi_S(\rho) = \text{Tr}_2 U\rho \otimes \mathbbm{1} U^{\dagger}
$$
Therefore, the Steinespring dilation theorem does not prevent us from checking if $\rho^{\Phi, S}$ is a partial trace of a Unitary map. In order to distinguish from the form of the dilation that appears in Stinespring's theorem, we refrain from using the term dilation and instead use \textit{parent}. We say that $\rho$ is a parent of $\rho^{\Phi, S}$ is $\rho$ acts on a product space and satisfies 
\begin{equation}
\rho^{\Phi, S} = \text{Tr}_2 \rho
\end{equation}
The question is, does $\rho^{\Phi, S}$ have a \textit{unitary parent}?

A unitary parent, among other things is also pure, i.e., $\rho=\ket{\psi}\bra{\psi}$ would be a rank-1 operator. Consider the spectral decomposition of $\rho^{\Phi, S}$:
\begin{equation}
\rho^{\Phi, S} = \sum_i \lambda_i \ket{\alpha_i}\bra{\alpha_i}
\end{equation}
Every pure parent $\rho$ would be of the form:
\begin{equation}
\rho = \left(\sum_i \sqrt{\lambda_i} \ket{\alpha_i}\otimes \ket{\beta_i}\right)\left(\sum_i \sqrt{\lambda_i} \bra{\alpha_i}\otimes \bra{\beta_i}\right)
\end{equation}
Here, $\ket{\beta_i}$ are vectors in dimension $(d')^2$ for some $d'$ and satisfy $\ket{\beta_i}\beta_j\rangle =\delta_{ij}$. One can find an arbitrary number of $\rho$'s that take this form. However, the question is, does there exist one of them that represents a unitary map? Unitarity implies that $\rho$ is maximally entangled between the input and the output qubits. To put it in a mathematical form, let us reshape the vectors $\ket{\alpha_i}$ and $\ket{\beta_i}$ into square matrices $\hat{\alpha}_i$ and $\hat{\beta}_i$ of size $d\times d$ and $d' \times d'$ respectively.  They satisfy
\begin{equation}
\begin{split}
\text{Tr}(\hat{\alpha}_i^{\dagger} \hat{\alpha}_j)&=\text{Tr}(\hat{\alpha}_i \hat{\alpha}_j^{\dagger})=\delta_{ij}\\
\text{Tr}(\hat{\beta}_i^{\dagger} \hat{\beta}_j)&=\text{Tr}(\hat{\beta}_i \hat{\beta}_j^{\dagger})=\delta_{ij}\\
\sum_{i, j} \sqrt{\lambda_i \lambda_j} &\hat{\alpha}_i^{\dagger} \hat{\alpha}_j \otimes \hat{\beta}_i \hat{\beta}_j^{\dagger} =\mathbbm{1}\\
\end{split}
\end{equation}
The last condition comes from the  unitary $U=\sum\sqrt{\lambda_i}\hat{\alpha}_i \otimes \hat{\beta}_i $. 
Thus, we have the following lemma:

\noindent\textbf{Lemma 3:} A doubly stochastic Choi matrix $\rho^{\Phi, S} = \sum_i \lambda_i \ket{\alpha_i}\bra{\alpha_i}$ has a unitary parent if and only if there exist square matrices $\hat{\beta}_i$ satisfying the conditions:
\begin{equation}
\begin{split}
\text{Tr}(\hat{\beta}_i^{\dagger} \hat{\beta}_j)&=\text{Tr}(\hat{\beta}_i \hat{\beta}_j^{\dagger})=\delta_{ij}\\
\sum_i \lambda_i \hat{\beta}_i\hat{\beta}_i^{\dagger}& =\sum_i \lambda_i \hat{\beta}_i^{\dagger} \hat{\beta}_i=\mathbbm{1}\\ 
\sum_{i, j} \sqrt{\lambda_i \lambda_j} &\hat{\alpha}_i^{\dagger} \hat{\alpha}_j \otimes \hat{\beta}_i \hat{\beta}_j^{\dagger} =\mathbbm{1}\\
\end{split}
\end{equation}

This is a necessary and sufficient condition. However, it is non-trivial to verify.  In theorem $2$ below, we provide a verifiable necessary condition.
\begin{figure*}
\includegraphics[scale=1]{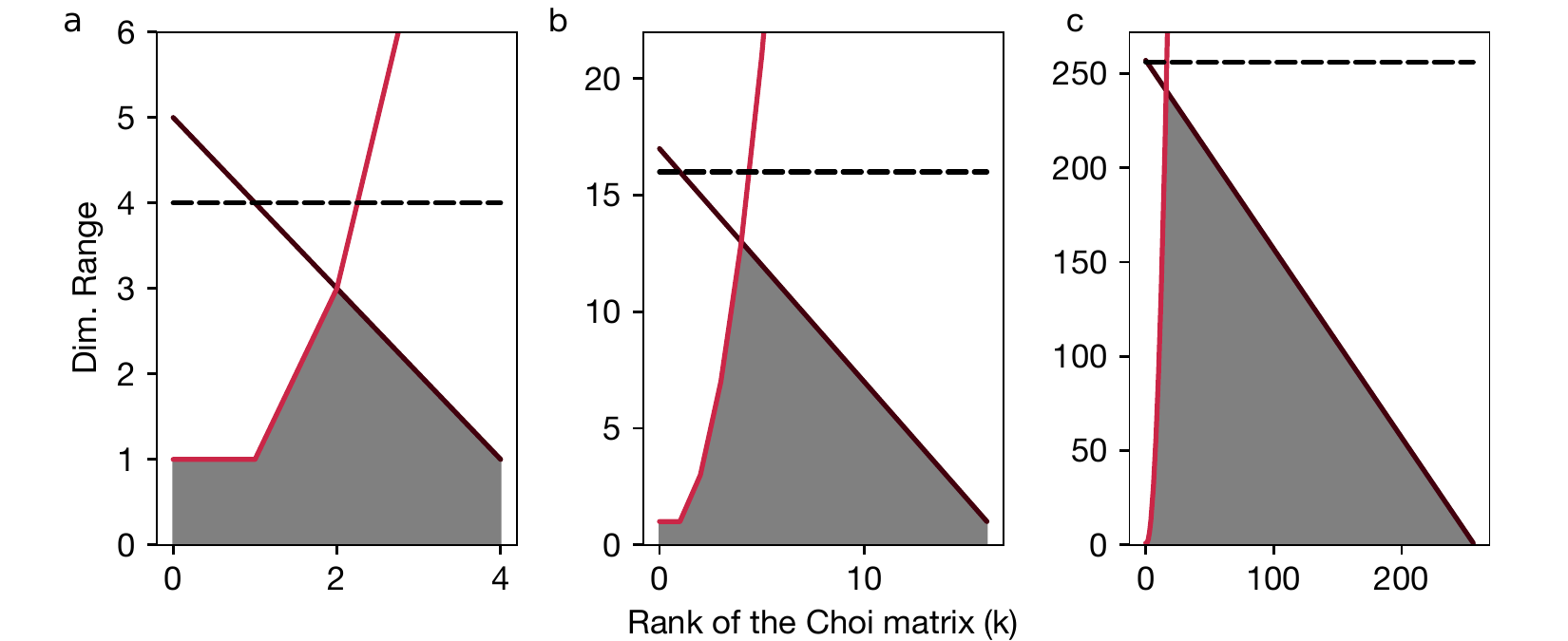}
\caption{\textbf{The rank property} \textbf{a}, \textbf{b} and \textbf{c} show the bounds $d^2-k+1$(black) and $k^2-k+1$ (red) from theorem 2, for $d=2, 4$ and $16$ respectively. The grey area consists of points that satisfy the inequalities in Eq.~\ref{rank_property}.  The black dashed line represents the limit $d^2$ to the dim. range of $\{\hat{\alpha}_i\hat{\alpha}_j^{\dagger}\} $. }\label{Fig4}
\end{figure*}

\noindent\textbf{Theorem 2: }A doubly stochastic Choi matrix $\rho^{\Phi, S} = \sum_{i=1}^k \lambda_i \ket{\alpha_i}\bra{\alpha_i}$ has a unitary parent \textit{only if}
\begin{equation}\label{rank_property}
\text{dim.  range}\{\hat{\alpha}_i\hat{\alpha}_j^{\dagger}\} \leq d^2-k+1, k^2-k+1
\end{equation}
Here, $k$ is the rank of $\rho^{\Phi, S} $.

We refer to Eq.~\ref{rank_property} as the \textit{rank property}, the violation of which would imply that there is no unitary parent and hence a non-unitary CP-indivisible error.  

\textbf{Proof: } We begin by understanding the implications of Lemma 3. The equation
\begin{equation}
\sum_{i, j} \sqrt{\lambda_i \lambda_j} \hat{\alpha}_i^{\dagger} \hat{\alpha}_j \otimes \hat{\beta}_i \hat{\beta}_j^{\dagger} =\mathbbm{1}
\end{equation}
can be re written in a different way:
\begin{equation}\label{unitarity2}
\sum_{i, j} \sqrt{\lambda_i \lambda_j} \left(\hat{\alpha}_i^{\dagger} \hat{\alpha}_j -\frac{1}{d}\delta_{ij}\mathbbm{1}\right)\otimes \left(\hat{\beta}_i \hat{\beta}_j^{\dagger}-\frac{1}{d}\delta_{ij}\mathbbm{1} \right)=0
\end{equation}
This follows from the double stochasticity conditions
$$
\sum_i \lambda_i \hat{\beta}_i\hat{\beta}_i^{\dagger}=\sum_i \lambda_i \hat{\beta}_i^{\dagger} \hat{\beta}_i=\mathbbm{1}
$$ 
and
$$
\sum_i \lambda_i \hat{\alpha}_i\hat{\alpha}_i^{\dagger} =\sum_i \lambda_i \hat{\alpha}_i^{\dagger} \hat{\alpha}_i=\mathbbm{1}
$$ 
Eq.~\ref{unitarity2} has strong implications on the range of $\{\hat{\alpha}_i^{\dagger}\hat{\alpha}_j\}$.  This is an expression of the form $\sum_i \mathbf{a}_i\otimes \mathbf{b}_i=0$ where $\{\mathbf{a}_i\}$ and $\{\mathbf{b}_i\}$ are vectors. This equation implies that the range space of  $\{\mathbf{b}_i\}$  is orthogonal to all of $\mathbf{a}_i$. To see this, le us re-write Eq.~\ref{unitarity2} in a matrix form. Let us define $k^2\times d^2 $ matrices $A$ and $B$:
\begin{equation}
\begin{split}
A &=[f(\hat{\alpha}_1^{\dagger}\hat{\alpha}_1),f( \hat{\alpha}_1^{\dagger}\hat{\alpha}_2), \cdots, f(\hat{\alpha}_k^{\dagger}\hat{\alpha}_k)]\\
B&=[f(\hat{\beta}_1^{\dagger}\hat{\beta}_1), f(\hat{\beta}_1^{\dagger}\hat{\beta}_2), \cdots, f(\hat{\beta}_k^{\dagger}\hat{\beta}_k)]\\
\end{split}
\end{equation}
Here, $f(\hat{\alpha}_j^{\dagger}\hat{\alpha}_j)$ represents \textit{flattening} the matrix $\hat{\alpha}_i^{\dagger} \hat{\alpha}_j -\frac{1}{d}\delta_{ij}\mathbbm{1}$ into a vector.  Eq.~\ref{unitarity2} can be written as
$$
AB^{\dagger} = BA^{\dagger} = 0
$$
Thus, range($B^{\dagger}$) $\subset$ NullSpace($A$). Moreover, 
$$
\text{rank}(A) \leq \text{nullity}(B), \text{nullity}(B^{\dagger}). 
$$
It remains to find upper bounds on the nullity of $B$.  Let $x_1, \cdots , x_k$ be scalars and $v= (x_1^*x_1, x_1^*x_2, \cdots, x_k^*x_k)$. It follows that
$$
Bv= f((x_1\hat{\beta}_1+\cdots +x_k \hat{\beta}_k)^{\dagger}(x_1\hat{\beta}_1+\cdots +x_k \hat{\beta}_k))
$$
The matrices $\{\hat{\beta}_i\}$ are linearly independent and therefore, for any matrix $X\in \text{span}\{\hat{\beta}_i\}$, $Bv=f(XX^{\dagger})$ has a solution. Therefore, the range of $B$ contains every such $XX^{\dagger}$.  This puts a lower bound on the rank of $B$ -- it should be at least $k$. Thus, the nullities of $B$ and $B^{\dagger}$ are bounded above by $k^2-k$ and $d^2-k$. The dimension of range of $\{\hat{\alpha}_i\hat{\alpha}_j^{\dagger}\} $ is one more than the rank of $A$ (due to the $\mathbbm{1}$). Thus, the result follows $\blacksquare$

Next, we address the question of quantification of the non-unitary, CP-indivisible errors using the rank property. Fig.~\ref{Fig4} shows that at large $k$, the bound set by the rank property increases with decreasing $k$. Therefore, an obvious choice is to remove the smallest eigenvalues of $\rho^{S, \Phi}$, until it satisfies the rank property.  If $\rho^{S, \Phi}$ satisfies the rank property after removing the smallest $\mu$ eigenvalues, the residue is the sum of the squares of these eigenvalues, which is an error estimate.

Finally, we address the question of sufficiency of the rank property.  In general, the above condition is not sufficient for the existence of a unitary parent.  Therefore, there can be non-unitary, CP-indivisible errors that dont violate the rank property.  However, similar to non-unitary, CP-divisible errors, one can show that physically reasonable CP-indivisible errors always lead to a violation of the rank property.  The shot-to-shot fluctuations of the unitary result in an increase in the rank of the Choi matrix.  Fig.~\ref{Fig4} shows that the bounds of the rank property are very stringent at high values of the rank of the Choi matrix.  Thus, for most physical shot-to-shot fluctuations, one can expect that the rank property will be violated. \\

\tocless\section{Unitary errors}\label{Unitary}
The above techniques don't work for unitary errors.  Consider, for instance, a target multi-qubit gate $\Phi_U(\rho)=U\rho U^{\dagger}$ and a real gate, $\Phi_{U'}(\rho)=U'\rho U^{'\dagger}$, which is also unitary, albeit a slightly different one.  That is, $||U^{\dagger}U'-\mathbbm{1}||_2$ is small.  The reduced Choi matrix of the real gate corresponding to a subset $S$,  $\Phi^{S, U'}$, while not unitary, satisfies both double stochasticity and the ran property. Moreover, it \textit{has} a unitary parent.  Therefore, no tests of unitary parent will detect the error. This error, which we refer to as unitary error, in general, can't be detected, unless some information is given about $U$.  The most common form of known information about $U$ is an observable or a set of observables that commute with it.

We discuss a few examples before proceeding further. If the multi-qubit gate is generated by a constant Hamiltonian, i.e.,  $U=e^{-iHt}$, then, observables of the form $X=H^{n}$ are conserved. In most practical cases, $H$ is a nearest neighbor or a next nearest neighbor Hamiltonian and is, therefore,  easy to measure using reduced Choi matrices corresponding to connected two-qubit subsets.  In general, if $H$ consists of at most $m-$qubit correlators, then $\langle H\rangle$ can be measured using reduced Choi matrices $\rho^{U', S}$ corresponding to all the $m-$qubit subsets $S$.  Besides $H$, many of the popular choices of $H$ have known symmetries, such as the total magnetization $X=\sum_i \sigma_{zi}$, in a Heisenberg chain.  Recently, a class of many-body Hamiltonians that feature \textit{Hilbert space fragmentation}, which have a large number of conserved observables have been studied~\cite{PhysRevLett.130.010201}.  A random unitary error would violate the conservation of such observables. 

For simplicity, let us assume that a conserved quantity $X=\sum_i X_i$ is known, which can be written as a sum of single qubit operators. What follows cam be generalized easily to other observables. The error in the expectation value of this observable is given by 
\begin{equation}
\begin{split}
&\sum_i \text{Tr}[\Phi^{U, \{i\}}(\rho_i)X_i-\Phi^{U', \{i\}}(\rho_i)X_i] \\
&= \sum_i \text{Tr}[(\rho_i X_i)-\Phi^{U', \{i\}}(\rho_i)X_i]
\end{split}
\end{equation}
Note that $\Phi^{\{i\}, U'}$ represents the reduced process corresponding to the subset $S=\{i\}$. The R.H.S of the above equation can be computed after measuring all the Choi matrices. It follows that 
\begin{equation}\label{Unitary_ineq}
\begin{split}
&\sum_i \sigma_{\max}(\Phi^{U, \{i\}}-\Phi^{U', \{i\}})||X_i||_2 \\
&\geq  \sum_i \text{Tr}[(\rho_i X_i)-\Phi^{U', \{i\}}(\rho_i)X_i]
\end{split}
\end{equation}
Thus, the most natural measure of unitary errors is given by 
\begin{equation}
\max_{\{\rho_i\}}| \sum_i \text{Tr}[(\rho_i X_i)-\Phi^{U', \{i\}}(\rho_i)X_i]|
\end{equation}
The maximization can be performed classically, after extracting the Choi matrices. This quantity is a lower bound on the sum of the maximum singular values in Eq.~\ref{Unitary_ineq}. Moreover, it misses out on the special unitary errors that maintain the commutation with $X$. However, statistically, this estimates the error within a factor of $O(1)$. \\

\tocless\section{Conclusion}\label{conclusions}

In the previous paper~\cite{https://doi.org/10.48550/arxiv.2210.04330}, we developed quantum metrology-based techniques to measure the reduced Choi matrix of a multi-qubit gate acting on $N$ qubits.  In this paper, we have followed it up with the computational aspects of using the reduced Choi matrix to benchmark the multi-qubit gate. In particular, we showed a number of properties of the reduced Choi matrix which can be computed efficiently on a classical computer and therefore can be used to benchmark it. They include double stochasticity, discussed in section~\ref{NU_markovian}, the rank property, developed in section~\ref{NU_Nmarkovian}, and symmetry properties, discussed in section~\ref{Unitary}. These properties represent a very small part of the information contained in the quantum process and therefore, they cannot characterize all \textit{mathematically plausible} errors.  However, we show that under certain \textit{physically reasonable} assumptions on the source of the errors, these properties can detect most of the errors.   For instance,  if the quantum computer is coupled to a bath, we can assume that the bath is in an ergodic state with a well-defined, finite temperature.  Interestingly, violation of the three properties represents errors with three different physical origins.  Therefore, these properties form a powerful set of tools to benchmark multi-qubit gates. 

The results in this and the previous paper are aimed at benchmarking quantum computers and simulators based on neutral atoms trapped in a tweezer array.  The metrological protocols described in the previous paper and the types of Hamiltonians and errors considered in this paper are suitable for such hardware systems. Therefore, we hope to be able to use these techniques to benchmark some of the useful multi-qubit gates in ultracold atomic systems in the near future.

Another possible application of our results is in quantum error mitigation~\cite{https://doi.org/10.48550/arxiv.2210.00921, PRXQuantum.3.040313}. The properties of Choi matrices identified in this paper can also be used to mitigate the error in multi-qubit gates. For instance, if an experimentally measured Choi matrix is not doubly stochastic, one can consider mitigating this error, either by finding the nearest doubly stochastic matrix to the measured one or by developing a map between Choi matrices that corrects for the double stochasticity violation. This would be an error mitigation strategy. One can also develop such strategies for the other two properties developed in this paper.

\paragraph*{\textbf{Acknowledgments}}
I thank Monika Aidelsburger for illuminating ``hallway" discussions.  I also thank Lukasz Cincio and Marco Cerezo for fruitful discussions. This work was supported by the European Union and  Deutsche Forschungsgemeinschaft (DFG, German Research Foundation) under Germany's Excellence Strategy -- EXC- 2111 -- 39081486. The work at LMU was additionally supported by DIP. This project has received funding from the European Union's Horizon 2020 research and innovation programme under the Marie Skłodowska-Curie grant agreement No 893181.

\paragraph*{\textbf{Competing interests}} The authors declare no competing interests. 

\bibliography{References}

\begin{thebibliography}{38}
\bibitem [{not()}]{note_metric}%
  \BibitemOpen
  \href@noop {} {}\bibinfo {note} {If $f(\rho, \rho')$ and $g(\rho, \rho')$ are
  two metrics satisfying $f\leq c_1 g$ and $g\leq c_2 f$, for constants $c_1$
  and $c_2$, then any rate of convergence proved using $f$ also holds with $g$,
  albeit with a different coefficient.}\BibitemShut {Stop}%
\end{thebibliography}


\appendix
\cleardoublepage

\setcounter{figure}{0}
\setcounter{page}{1}
\setcounter{equation}{0}
\setcounter{section}{0}

\renewcommand{\thepage}{S\arabic{page}}
\renewcommand{\thesection}{S\arabic{section}}
\renewcommand{\theequation}{S\arabic{equation}}
\renewcommand{\thefigure}{S\arabic{figure}}
\onecolumngrid
\begin{center}
\huge{Supplementary Information}
\vspace{5mm}
\end{center}
\twocolumngrid
\normalsize
\tableofcontents

\section{Quantification of errors in the Choi matrix}
In this section, we discuss the various metrics of the error in a Choi matrix and explain why the most appropriate choice is the maximum singular value, used in the main text.  Let $\Phi$ and $\Phi'$ be two CP maps and $\rho^{\Phi}, \rho^{\Phi'}$  are the corresponding Choi matrices.  In the previous paper~\cite{https://doi.org/10.48550/arxiv.2210.04330}, we used the Schatten-2 norm, $||\rho^{\Phi}-\rho^{\Phi'}||_2 $ to study the convergence rate in the reduced process tomography.  In this case, the precise choice of the metric is unimportant, since the rate of convergence remains the same for all of them~\cite{note_metric} . However, the purpose of the metric in this paper is to develop benchmarks and therefore, we delve deeper into the question of the most appropriate measure. 

In benchmarking, the relevant quantity is the error in the expectation value of an observable. That is, as mentioned in the main text, $\Delta \langle O \rangle_\rho =\langle O\rangle_{\Phi(\rho)} - \langle O\rangle_{\Phi'(\rho)} = \text{Tr}(\epsilon O\otimes \rho)$, where $\epsilon= \rho^{\Phi}-\rho^{\Phi'}$.  Accordingly,  we define $\sigma_{\max}(\Phi-\Phi')$ as the metric of the error.  Before discussing the properties of this metric in relation to the Schatten-2 norm, we clarify a crucial difference between the singular values of the matrix $\rho^{\Phi}-\rho^{\Phi'}$ and the singular values of the map $\Phi-\Phi'$. 

The matrix $\rho^{\Phi}$ can be interpreted also as a map $\mathbbm{C}^d\otimes \mathbbm{C}^d \rightarrow \mathbbm{C}^d\otimes \mathbbm{C}^d$, besides representing the map $\Phi : \mathbbm{C}^{d\times d}\rightarrow \mathbbm{C}^{d\times d}$.  We illustrate the difference between these two maps. Let $\{\ket{1}, \cdots, \ket{d}\}$ be a basis.  Let $\rho = \sum \rho_{ij}\ket{i}\bra{j} \in \mathbbm{C}^{d\times d}$ and $\ket{\psi} =\sum \psi_{ij}\ket{i}\otimes\ket{j}\in \mathbbm{C}^d\otimes \mathbbm{C}^d$.  Their images under the two maps are
\begin{equation*}
\begin{split}
\Phi(\rho) &= \sum \rho_{ij}\Phi(\ket{i}\bra{j}) = \sum \rho^{\Phi}_{ik; jl}\rho_{ij}\ket{k}\bra{l}\\
\rho^{\Phi}\ket{\psi} &= \sum \rho^{\Phi}_{ik; jl}\psi_{jl}\ket{i}\otimes \ket{k}
\end{split}
\end{equation*}
Similarly, $\epsilon$ has two interpretations. The singular values corresponding to the two interpretations can be quite different.  It is straightforward to show that the sum of the squares of the singular values of the two maps are equal. Indeed,
\begin{equation}\label{soss_1}
\sum\sigma^2(\Phi) = \sum_{i, j}||\Phi(\ket{i}\bra{j})||^2 = \sum_{ijkl}|\rho^{\Phi}_{ik; jl}|^2
\end{equation} 
and
\begin{equation}\label{soss_2}
\sum\sigma^2(\rho^{\Phi}) =\sum_{ j, l}||\rho^{\Phi}\ket{j}\otimes \ket{l}||^2 = \sum_{ijkl}|\rho^{\Phi}_{ik; jl}|^2
\end{equation} 
However, the individual singular values can be very different. We provide two extreme examples:

\noindent\textit{Example 1: } Consider the identity, $\Phi(\rho)=\rho$. The sum of the squares of the singular values is $d^2$. The map $\Phi$ has $d^2$ singular values, all equal to $1$. The matrix $\rho^{\Phi}$ has $d^2$ singular values, one of which is equal to $d$ and the rest are $0$. Note that $\rho^{\Phi}$ is a rank-1 matrix. 

\noindent\textit{Example-2: } Consider the depolarizing map $\Phi(\rho)=\frac{1}{d}\mathbbm{1}_d \forall \rho$. In this case, $\rho^{\Phi} =\frac{1}{d}\mathbbm{1}_{d^2} $. The sum of the squares of the singular values is $1$.  The map $\Phi$ has one singular value equal to $1$ and the rest are zero. The matrix $\rho^{\Phi}$ has all of its singular values equal to $\frac{1}{d}$. 

It is now clear from Eqs.~\ref{soss_1}\&\ref{soss_2} that 
\begin{equation}
\sum \sigma^2(\Phi-\Phi') = ||\rho^{\Phi}-\rho^{\Phi'}||_2^2
\end{equation}
Thus, we obtain the relation between the two metrics:
\begin{equation}\label{upper}
\sigma_{\max}(\Phi-\Phi')\leq  ||\rho^{\Phi}-\rho^{\Phi'}||_2
\end{equation}
and
\begin{equation}\label{lower}
\sigma_{\max}(\Phi-\Phi')\geq \frac{1}{d} ||\rho^{\Phi}-\rho^{\Phi'}||_2
\end{equation}
The last inequality follows from $ \sigma_{\max}^2(\Phi-\Phi') \geq  \sigma^2(\Phi-\Phi')$. The above two inequalities show that the convergence rates proved in ref.~\cite{https://doi.org/10.48550/arxiv.2210.04330} also apply to the metric $\sigma_{\max}(\Phi-\Phi')$. 

We now consider a specific example to compute this metric and compare it with the Schatten norm.  Let $\Phi_{id}: \rho \mapsto \rho$ be the identity map for a single qubit, i.e., $d=2$ and $\Phi_{BR}$ be the Bloch-Redfield decay. The corresponding Choi matrices are

\begin{equation}
\rho^{\Phi_{id}}  =\left(
\begin{array}{cc|cc}
1 &  0 &  0 & 1\\
0 &  0 &  0 & 0\\
\hline
0 &  0 &  0 & 0\\
1 &  0 &  0 & 1\\
\end{array}
\right)
\end{equation}
\begin{equation}
\rho^{\Phi_{BR}}=\left(
\begin{array}{cc|cc}
1 &  0 &  0 & e^{-\Gamma_2 t}\\
0 &  0 &  0 & 0\\
\hline
0 &  0 &  1-e^{-\Gamma_1 t} & 0\\
e^{-\Gamma_2 t} &  0 &  0 & e^{-\Gamma_1 t}\\
\end{array}
\right)
\end{equation}
The difference is 
\begin{equation}
\rho^{\Phi_{BR}}-\rho^{\Phi_{id}}=\left(
\begin{array}{cc|cc}
0 &  0 &  0 & e^{-\Gamma_2 t}-1\\
0 &  0 &  0 & 0\\
\hline
0 &  0 &  1-e^{-\Gamma_1 t} & 0\\
e^{-\Gamma_2 t}-1 &  0 &  0 & e^{-\Gamma_1 t}-1\\
\end{array}
\right)
\end{equation}
It follows that 
\begin{equation}
\begin{split}
\sigma_{\max}(\Phi_{id}-\Phi_{BR}) &= \max\{\sqrt{2}(1-e^{-\Gamma_1 t}), 1-e^{-\Gamma_2 t}\}\\
||\rho^{\Phi_{id}}-\rho^{\Phi_{BR}}||_2 &= \sqrt{2(1-e^{-\Gamma_1 t})^2,+ (1-e^{-\Gamma_2 t})^2}\\
\end{split}
\end{equation}

\begin{figure}
\includegraphics[scale=1]{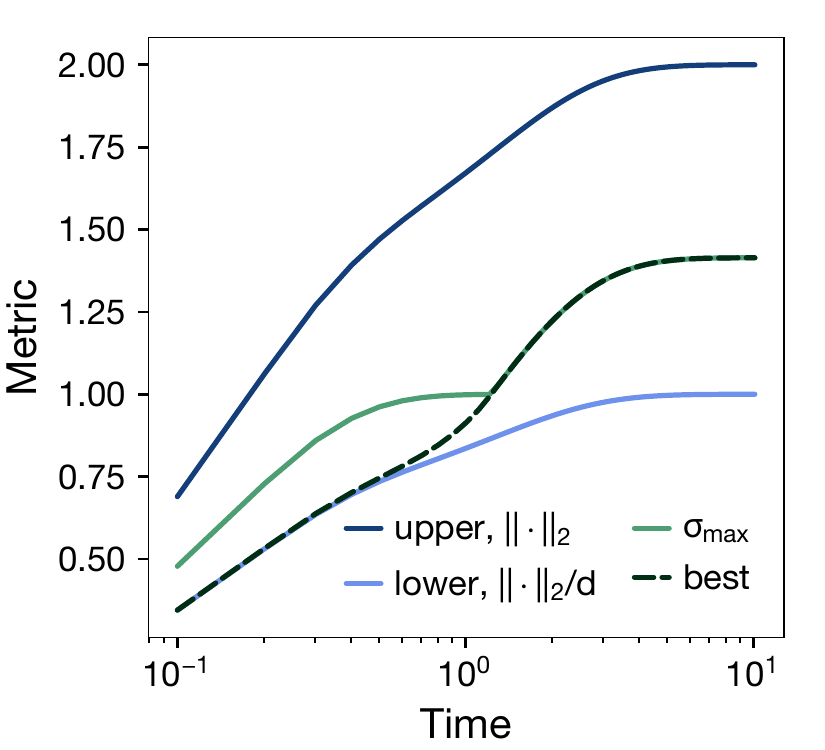}
\caption{\textbf{Measures of error in the Choi matrix: } A comparision of $\sigma_{\max}$ with the upper and lower bounds, shown in Eq.~\ref{upper} and Eq.~\ref{lower} respectively, for the Bloch-Redfield model. The black dashed line is the exact error, given by $\max_{\ket{\psi}}||\Phi(\ket{\psi}\bra{\psi})-\Phi'(\ket{\psi}\bra{\psi})||_2$. }\label{S1}
\end{figure}

Finally, we discuss a possible improvement of the metric, beyond $\sigma_{\max}$.  Indeed, one can consider $\max_{\rho}||\Phi(\rho)-\Phi'(\rho)||_2$. By convexity of the $2-$norm, this is equivalent to $\max_{\ket{\psi}}||\Phi(\ket{\psi}\bra{\psi})-\Phi'(\ket{\psi}\bra{\psi})||_2$. This is in fact a constrained optimization of a degree-4 polynomial, which can be done using semi-definite programming. However,  it is much more complicated and in general, it is not likely to be much different from $\sigma_{\max}$.  Fig.~\ref{S1} shows that this measure is quite close to $\sigma_{\max}$ for the case of the Bloch-Redfield model.

\section{CP-divisibility}\label{CP_divisibility}

In this section, we discuss a few examples illustrating the concept of divisibility of CP maps.  Recall that a trajectory of a map $\Phi_t$ is divisible iff
\begin{equation}
\Phi_{t}\circ\Phi_{t'} = \Phi_{t+t'}
\end{equation}
We provide a physically motivated example of a divisible (i.e., Markovian) and an indivisible (i.e., non-Markovian) process. 

\noindent \textbf{Example 1: Markovian} Let us consider the Bloch-Redfield decay of a single qubit. The Choi matrix is 

\begin{equation}
\rho^{\Phi_{BR}(t)}=\left(
\begin{array}{cc|cc}
1 &  0 &  0 & e^{-\Gamma_2 t}\\
0 &  0 &  0 & 0\\
\hline
0 &  0 &  1-e^{-\Gamma_1 t} & 0\\
e^{-\Gamma_2 t} &  0 &  0 & e^{-\Gamma_1 t}\\
\end{array}
\right)
\end{equation}
It follows that 
\begin{equation}
\begin{split}
\rho^{\Phi_{BR}(t)\circ \Phi_{BR}(t')}=&\left(
\begin{array}{cc|cc}
1 &  0 &  0 & e^{-\Gamma_2 (t+t')}\\
0 &  0 &  0 & 0\\
\hline
0 &  0 &  1-e^{-\Gamma_1 (t+t')} & 0\\
e^{-\Gamma_2 (t+t')} &  0 &  0 & e^{-\Gamma_1 (t+t')}\\
\end{array}
\right) \\
=& \rho^{\Phi_{BR}(t+t')}
\end{split}
\end{equation}
Thus, this process is the solution to a Lindblad master equation. In fact, in the extreme case, $\Gamma_2=\Gamma_1/2$, this process results from coupling to a thermal bath at zero temperature (Wigner-Weisskopf theory). When $\Gamma_1=0$, it results from coupling to a thermal bath at infinite temperature. 

\noindent \textbf{Example 2: non-Markovian:} Let us consider a model of dephasing due to Gaussian fluctuations, in a spin-1/2 system. The energy gap between the spin states $\ket{\uparrow}$ and $\ket{\downarrow}$ is given by $\omega=-\frac{\gamma_B}{2}B$, where $B$ is the applied magnetic field and $\gamma_B$ is the gyro-magnetic ratio.  The Choi matrix representing time evolution of a single spin is
\begin{equation}
\rho^{\Phi(t)}(B) = \left(
\begin{array}{cc|cc}
1 &  0 &  0 & e^{-i\omega t}\\
0 &  0 &  0 & 0\\
\hline
0 &  0 & 0 & 0\\
e^{i\omega t} &  0 &  0 & 1\\
\end{array}
\right)
\end{equation}
Let us consider an  ensemble of such spins, with a magnetic field modelled by a normal distribution: $B\sim \frac{1}{\sqrt{2\pi}\sigma_B}e^{-\frac{(B-B_0)^2}{2\sigma_B^2}}$.  The resulting Chi matrix is 

\begin{equation}
\begin{split}
 \rho^{\Phi(t)}&=\int dB \frac{1}{\sqrt{2\pi}\sigma_B}e^{-\frac{(B-B_0)^2}{2\sigma_B^2}} \rho^{\Phi(t)}(B) \\
 &=\left(
\begin{array}{cc|cc}
1 &  0 &  0 & e^{-i\omega_0 t -\sigma^2 t^2}\\
0 &  0 &  0 & 0\\
\hline
0 &  0 &  0 & 0\\
e^{i\omega_0 t -\sigma^2 t^2}&  0 &  0 & 1\\
\end{array}
\right)
\end{split}
\end{equation}
Here, $\omega_0 = -\frac{\gamma_B}{2}B_0$ and $\sigma= \frac{1}{\gamma_B\sigma_B}$.  It follows that 
\begin{equation}
\begin{split}
 \rho^{\Phi(t)\circ\Phi(t')} &=\left(
\begin{array}{cc|cc}
1 &  0 &  0 & e^{-i\omega_0 (t+t') -\sigma^2 (t^2+t'^2)}\\
0 &  0 &  0 & 0\\
\hline
0 &  0 &  0 & 0\\
 e^{i\omega_0 (t+t') -\sigma^2 (t^2+t'^2)}&  0 &  0 & 1\\
\end{array}
\right)\\
&\neq \rho^{\Phi(t+t')}
\end{split}
\end{equation}
Note that the Gaussian, $e^{-\sigma^2 t^2}$ does not multiply like the exponential map, $e^{-\Gamma t}$.   This shows that there is no Lindblad master equation that produces this trajectory of Choi matrices, from GKLS theorem.  The physical intuition is, in order to continue the propagation starting from some time $t$, one needs to either know the states of all of the members of the ensemble or the density matrices at each time before $t$ --- knowing the state at $t$ is insufficient to propagate it further under this map. 


\end{document}